\documentclass[letterpaper]{article} 
\usepackage{aaai25}  
\usepackage{times}  
\usepackage{helvet}  
\usepackage{courier}  
\usepackage[hyphens]{url}  
\usepackage{graphicx} 
\usepackage{natbib}  
\usepackage{caption} 
\usepackage{algorithm}
\usepackage{algorithmic}

\usepackage{newfloat}
\usepackage{listings}
\DeclareCaptionStyle{ruled}{labelfont=normalfont,labelsep=colon,strut=off} 
\floatstyle{ruled}
\newfloat{listing}{tb}{lst}{}
\floatname{listing}{Listing}
\title{Learning About Algorithm Auditing in Five Steps: Scaffolding How High School Youth Can Systematically and Critically Evaluate Machine Learning Applications}

\author {
    Luis Morales-Navarro\textsuperscript{\rm 1},
    Yasmin B. Kafai\textsuperscript{\rm 1},
    Lauren Vogelstein\textsuperscript{\rm 2},
    Evelyn Yu\textsuperscript{\rm 1},
    Danaë Metaxa\textsuperscript{\rm 1}
}
\affiliations {
    \textsuperscript{\rm 1}University of Pennsylvania\\
    \textsuperscript{\rm 2}Teachers College, Columbia University \\
    luismn@upenn.edu, kafai@upenn.edu, lev2124@tc.columbia.edu, evelynyu@upenn.edu, metaxa@upenn.edu
}

\usepackage{bibentry}

\begin{document}

\maketitle
\frenchspacing
\begin{abstract}
While there is widespread interest in supporting young people to critically evaluate machine learning-powered systems, there is little research on how we can support them in inquiring about how these systems work and what their limitations and implications may be. Outside of K-12 education, an effective strategy in evaluating black-boxed systems is algorithm auditing—a method for understanding algorithmic systems’ opaque inner workings and external impacts from the outside in. In this paper, we review how expert researchers conduct algorithm audits and how end users engage in auditing practices to propose five steps that, when incorporated into learning activities, can support young people in auditing algorithms. We present a case study of a team of teenagers engaging with each step during an out-of-school workshop in which they audited peer-designed generative AI TikTok filters. We discuss the kind of scaffolds we provided to support youth in algorithm auditing and directions and challenges for integrating algorithm auditing into classroom activities. This paper contributes: (a) a conceptualization of five steps to scaffold algorithm auditing learning activities, and (b) examples of how youth engaged with each step during our pilot study.
\end{abstract}

%

\section{Introduction}
Given the growing presence of artificial intelligence (AI) and machine learning (ML) in young people’s lives, it is critical to provide them with the resources they need to engage with, create, and evaluate AI/ML applications. AI education efforts have focused on developing AI/ML literacy so that young people can interact with AI/ML technologies as critical users, designers, and evaluators \cite{dennison2024consumers, long2020ai, touretzky2023machine}. The lack of transparency in ML models adds an additional barrier to understanding and critically engaging with AI/ML concepts. Furthermore, existing research pays little attention to assisting youth in investigating the limitations and implications of AI/ML systems \cite{van2023emerging}, focusing instead on discussing issues and providing direct instruction \cite{morales2024unpacking}, with little emphasis on how youth can empirically and critically assess system behaviors.

In algorithmic accountability and human-centered computing research, algorithm auditing has become an effective strategy for investigating and comprehending how AI/ML systems behave. Algorithm auditing involves "repeatedly querying an algorithm and observing its output in order to draw conclusions about the algorithm's opaque inner workings and possible external impact" \cite{metaxa2021auditing}. The majority of algorithm audits have been conducted either by experts or adult end-users to identify potential problematic system behaviors in AI/ML-powered systems \cite{bandy2021problematic, lam2023sociotechnical}. More recently, this research has been expanded to examine how  high school youth participate in auditing practices \cite{solyst2023potential, morales2024youth, walker2022liberatory}. These studies are promising in demonstrating how auditing activities can assist youth in critically evaluating and understanding AI/ML systems, as well as making connections to their lived experiences as users and designers. However, a common challenge they present is how to scaffold activities so that auditing moves from one-off observations to more systematic evaluations.

In this paper, we address this issue by conceptualizing a five-step auditing process that youth can engage with in learning activities to critically evaluate AI/ML systems. We base our five steps on common auditing methods used by expert auditors \cite{metaxa2021auditing, bandy2021problematic} and research on user-driven auditing practices \cite{devos2022toward, shen2021everyday, lam2023sociotechnical}. We conducted a two-week workshop with 16 youth (ages 14–15) in which they designed, tested, and audited TikTok filters. In this paper, we address the following research question: How can youth engage with each step of the auditing process while auditing peer-designed filters? We analyzed youth-created artifacts and video and screen recordings of youth engagement with each step to create a case study. We discuss the kind of scaffolds we provided to support youth in each step of algorithm auditing and directions and challenges for integrating algorithm auditing into classroom activities.

\section{Background}
Over the past decade, K-12 education researchers have started turning their attention to the design and study of critical computing learning activities \cite{ko2020time}. These activities often involve critical inquiry by approaching computing applications as sociotechnical systems and having learners investigate how computing systems work and what their implications are \cite{morales2023conceptualizing}. Within critical inquiry, deconstruction involves evaluating and reflecting on the values and intentions embedded in sociotechnical systems and considering their possible impact on people and the environment \cite{dindler2020computational, schaper2022computational}. This entails: (1) evaluating a system's inputs, outputs, and materials (such as data); (2) researching how a system is actually used; (3) investigating the values, worldviews, and presumptions that are ingrained in the system; and (4) determining how a system may affect individual people, communities, and the environment \cite{dindler2020computational}. While efforts to engage learners in critical computing activities are becoming increasingly common, particularly in the case of ML education, instructional activities that address issues of justice and ethics tend to be based on discussion or direct instruction \cite{morales2024unpacking} without methodically evaluating how systems work.

One promising activity to engage learners in evaluating AI/ML systems and investigate potential harmful biases is algorithm auditing. As described above, algorithm auditing involves querying and observing the outputs of systems \cite{metaxa2021auditing}. Auditing is different from traditional testing and evaluation in various ways \cite{metaxa2021auditing}. Unlike other types of testing, auditing is systematic, with the goal of drawing system-wide conclusions rather than individual test case results. Audits are also generally external evaluations performed by independent third parties without insider access, based on externally measured system behaviors. Algorithm auditing has been used in rigorous studies conducted by experts on racial housing discrimination in algorithmic systems, employee recruitment applications, recommendation systems, AI/ML-powered diagnosis and care systems in healthcare, and search engines (for a review of expert algorithm audits, see Bandy \shortcite{bandy2021problematic}; for details on the method, see Metaxa et al., \shortcite{metaxa2021auditing}). Recent work in auditing has explored how systems’ end-users can, without auditing expertise, be engaged in identifying harmful algorithmic behaviors through auditing practices. In one paradigm, adult users collaborate with experts, who explicitly task them with conducting an algorithm audit and provide friendly user interfaces for doing so (see Lam et al., \shortcite{lam2023sociotechnical}). Meanwhile, other studies have examined the phenomenon of  bottom-up, user-driven auditing practices that emerge organically among some user bases in the absence of explicit instructions \cite{devos2022toward, shen2021everyday}. 

In K-12 AI education, several efforts have explored different approaches to auditing-related activities. For instance, Walker and colleagues \shortcite{walker2022liberatory} proposed having youth conduct “evocative audits” to research the positive and negative ways in which computing systems affect the socio-political realities of their communities. Solyst et al. \shortcite{solyst2023potential} adapted user-driven everyday auditing tasks, such as making observations of Google search results \cite{devos2022toward} into activities for middle schoolers, noting that there is plenty of potential for youth to be involved in identifying and mitigating harmful algorithmic biases. Morales-Navarro et al. \shortcite{morales2024youth} designed a workshop in which high school participants designed ML-powered physical computing projects and then audited their peers’ projects. In a clinical interview study, they found that after the workshop, participants showed more nuanced understandings of data issues that may cause harmful biases, reflecting on how potential harmful biases and behaviors may emerge from the qualities of the data used to train models. Collectively, these observations support the idea that youth-driven algorithm audits are feasible, connecting to young people’s everyday experiences while also deepening their understanding of algorithmic systems. However, the auditing activities observed in these studies mostly center on a single or a couple of observations without engaging youth in the systematic analysis of inputs and outputs and the full process of conducting an audit from beginning to end. Thus, a major challenge is to identify and articulate an auditing approach that is systematic, accessible to, and actionable by youth.

\subsection{Auditing Algorithmic Systems in Five Steps}
To address this challenge, we turn to the research on algorithm auditing and conceptualize five different steps that make up the process of auditing algorithms. We recognize that not all algorithm audits are conducted the same way \cite{bandy2021problematic, costanza2022audits}; these steps are a general blueprint of one common process of auditing algorithmic systems \cite{metaxa2021auditing}. Based on our review, algorithm auditing involves: (1) developing a hypothesis, (2) generating a set of systematic, thorough, and thoughtful inputs to test the hypothesis, (3) running the test, (4) analyzing the data, and (5) reporting the results. Before discussing our case study with youth, we will walk through each step using an illustrative example of an expert-driven audit entitled “An Image of Society” (hereafter: AIS), conducted by Metaxa and colleagues \shortcite{metaxa2021image}, in which they investigated the effect of gender and racial representation in image search results for common workforce occupations. We also draw on examples from other expert-driven and user-driven audits to demonstrate common approaches to each step. 

\subsubsection{Step 1: Developing a Hypothesis}

To begin an audit, researchers must first generate a hypothesis about the kind of behavior they expect the algorithmic system being audited to exhibit. In AIS, the hypothesis was that gender and racial biases would be reflected in Google search results. This hypothesis was developed as a follow-up to an earlier audit by Kay et al. \shortcite{kay2015unequal} that studied how different levels of gender diversity in search results affected users’ perceptions of occupations and the search engine. Hypotheses are often based on auditors’ personal experiences using AI/ML systems. This was the case, for example, in the Gender Shades study by Buolamwini \& Gebru \shortcite{buolamwini2018gender}, where Buolamwini’s personal experiences with facial recognition led the team to investigate racial disparities in gender classification systems. At the same time, in user-driven audits, everyday people come up with hypotheses based on observed unexpected behaviors \cite{devos2022toward} and are often collectively formulated \cite{shen2021everyday}. In one such high-profile user-driven audit example of Twitter’s image cropping algorithm, one Twitter user, through his ordinary use of the platform, on one instance noticed that when an image included the faces of two people, one with light skin and one with dark skin, the platform would crop the darker-skinned individual out of the image preview shown on Twitter’s main feed. This led the user to tweet the hypothesis that Twitter’s image cropping algorithm was racially biased. Hypothesis development underpins nearly all auditing efforts, whose various subsequent methods focus on providing evidence for or against that claim.

\subsubsection{Step 2: Generating a Set of Systematic, Thorough, and Thoughtful Inputs} 

In the second step of most expert audits, auditors design a set of inputs to the system in question, carefully selected so as to rigorously test their hypothesis. These are ideally systematic, thorough, and thoughtful, such that their results could constitute convincing proof of the hypothesis. In AIS, researchers generated a list of 100 occupations to test their hypothesis on a wide range of jobs, some with racial and gender parity and some with documented racial and gender inequities, to be able to observe if search results were actually biased or accurately reflected the real-world gender and racial make-up of those occupations. In user-driven audits, auditors may be less systematic, with end users’ planning being more ad hoc and reactive to system behaviors or their own personal experience \cite{devos2022toward}. However, thanks to the scaling power of social media, some user-driven audits can approach expert levels of systematicity. Clearly, unlike the generation of a hypothesis, this step may be more compromised in audits directly involving users. This has implications for learning activities with youth, as we discuss in our case study.

\subsubsection{Step 3 - Running a Test}

The next step, running the tests, involves repeatedly querying the algorithmic system using the inputs designed in the step above, and recording the system’s outputs. In AIS, researchers conducted search queries of Google Image Search  for each of the 100 selected occupations and collected the top 100 images for each search using automated web scrapers. Other expert-driven audits have been conducted using both manual and automated means. That was the case for Sweeney’s \shortcite{sweeney2013discrimination} study of racial discrimination in online ad delivery, in which she investigated whether a Google search for a Black-sounding name resulted in ads suggestive of an arrest record compared to White-sounding names, with the researcher conducting 1373 manual searches and 812 automated ones. In user-driven audits that occur organically, often testing is impromptu, for instance, in the Twitter case with users replicating tests conducted by other users. More recently, researchers have built tools for user-driven auditing that support users in running and documenting their tests \cite{lam2023sociotechnical}.

\subsubsection{Step 4 - Analyzing the Data}

After testing, data is analyzed by describing the findings and/or conducting statistical analyses. In the AIS study, researchers recruited crowd-workers through Amazon Mechanical Turk to label the race and gender of the image outputs from the search results, with the goal of understanding how outside viewers perceived the people in those images. Then they used statistical tests to compare gender representation in Google Image Search results to 2020 data on occupation demographics from the United States Bureau of Labor and Statistics (BLS), discovering evidence of men's systematic over-representation. In terms of race, they found stronger systematic over-representation of the dominant group, white people, relative to people of color. In other expert-driven studies, data analysis involves grouping outputs and reporting descriptive statistics, rather than conducting statistical hypothesis testing. In Sweeney’s audit, discussed above, for instance, she calculated the percentage of arrest ads and neutral ads delivered after searching for Black- and White-sounding names on Google. In some user-driven audits, analysis may be similarly descriptive or even qualitative, with users inspecting and interpreting data based on their own lived experiences \cite{devos2022toward}, or seeing if they can replicate the findings of other users \cite{shen2021everyday}. In short learning activities, students could analyze patterns in the outputs and, like Sweeney, calculate percentages. 

\subsubsection{Step 5 - Reporting the Results}

An important part of conducting an audit is to report its results in order to effect change. Expert-driven audits, such as AIS, often become academic papers that are presented at conferences like CSCW or FAccT, and can lead to discussions directly with the teams responsible for the product. Other expert-driven audits lead to long-form journalism articles \cite{nicoletti23} or interactive websites where people can explore the data and see the analysis and results \cite{bow18}, with the goal of raising public awareness. While some reports are directed toward informing future researchers, others target policymakers or the general public.   

Distilling auditing into this five-step model and adapting it for learning activities requires us to think about algorithm auditing from a learner-centered perspective \cite{soloway1994learner, hsi1998learner}. This involves carefully considering the kind of scaffoldings that learners may need, their diverse lived experiences, and the shift in goals (from auditing a system to learning about and from auditing) \cite{soloway1994learner}. As such, learning activities should incorporate aspects of expert-driven auditing to support learners in understanding how and why audits are conducted. They should also draw from user-driven auditing practices to incorporate and speak to youths’ everyday lived experiences as users of AI/ML systems. In the following sections, we share the iterative design story of supporting high school youth to audit to foreground the shifts in specificity in the scaffolds we developed and the increased amount of time participants needed to thoroughly engage in each of the five auditing steps. 

\section{Methods}

\begin{table*}[t]
\centering
\begin{tabular}{l|l|lll}
\textbf{Activity/Step}                      & \textbf{Length} & \textbf{Scaffolds}                                                                                                                          \\
\textit{Step 1: Developing a Hypothesis}    & 1.5 hours         & \begin{tabular}[c]{@{}l@{}}Provided examples of hypotheses, modeled how to generate a \\ hypothesis together, time for open-ended exploration or play.\end{tabular}  \\
\textit{Step 2: Generating a set of inputs} & 2 hours         & \begin{tabular}[c]{@{}l@{}}Designed a two-axis input organizer and asked the \\ youth to fill it out with at least 30 images.\end{tabular}                    \\
\textit{Step 3 - Running tests}             & 2 hours         & \begin{tabular}[c]{@{}l@{}}Designed a table for students to record input and output \\ pairs and take notes on specific things in their outputs.\end{tabular}  \\
\textit{Step 4 - Analyzing the data}        & 1 hour         & \begin{tabular}[c]{@{}l@{}}Suggested ideas for the kind of analysis the students could do \\ (calculating percentages/describing in detail the outputs).\end{tabular}  \\
\textit{Step 5 - Reporting results}         & 1 hour         & \begin{tabular}[c]{@{}l@{}}Recommended potential audiences and gave them \\ examples of audit reports done in previous workshops.\end{tabular}       
\end{tabular}
\caption{Auditing steps and scaffolds provided to youth.}
\label{table1}
\end{table*}

\subsection{Context \& Design}
As the context in which to study the affordances of algorithm auditing activities, we worked with high school youth enrolled in a four-year after-school program, Science Investigators (a pseudonym), at a science center in the Northeastern United States. We offered fall, spring, and summer workshops on the topic of AI/ML algorithm auditing as part of the Science Investigators  programming. As an iterative design-based research study \cite{brown1992design, barab2014design}, we designed and implemented three sets of workshops during which we were able to refine our approach to supporting high schoolers through algorithm auditing activities. In all of the work we conducted, we received informed consent from youth’s parents and assent from the youth themselves. The analysis in this paper focuses on the third iteration of this work, conducted over a two-week summer workshop with 14 rising sophomores (14-15 years old) in the Science Investigators program. We worked with six female youth, one non-binary student, and seven male youth. Participants largely came from marginalized racial backgrounds, all but one as either African American, Asian American, or multiracial. 

We developed the procedure in our third workshop based on the first two series, which were conducted the previous fall and spring. In our fall participatory design workshop with juniors and seniors (15-17 years old), we had learners informally evaluate generative AI TikTok filters that transform pictures taken by users into stylized images. In that workshop, we tasked participants to find moments in which these did not work as expected.  This procedure was based on and closely matched user-driven emergent auditing. We observed that some of the hypotheses and conclusions drawn by youth were ad hoc, and often their data collection was not methodical. In the second workshop series, conducted in the spring, we decided to introduce youth to the structure of formal expert-driven audits and adapt this structure with elements from user-driven audits. At the start of those workshops, we gave examples for the first four steps of algorithm auditing: “come up with a hypothesis, [...] carefully come up with a set of inputs, [...] run the test and measure the system’s outputs, [and lastly] analyze the those results to see if your hypothesis is supported.” This was the first time we, as a research team, began to think about algorithm auditing as being composed of sequential steps. We added a fifth step to the process — creating and sharing an audit report — to emphasize the social motivation for conducting audits. During the spring workshops, we piloted using these five steps to scaffold learning activities in which youth audited generative AI anime TikTok filters.

In the summer workshop, we met with youth for 3.5 hours each day and had two full weeks to continue working with the same cohort (now rising sophomores) from the spring workshop. In the first week, youth designed filters and audited each other’s filters. In the second week, we shifted focus to auditing the tool used to design filters. In this paper, we focus on the first week’s activities in which participants designed TikTok filters using TikTok’s Effect House, a software to create filters, and audited each other's filters. 

With more time, we were able to better scaffold each step of the auditing process compared to the spring (see Table \ref{table1}). In Step 1, we gave participants time for open-ended exploration with a filter their peers had designed with the intention of paying attention to potentially problematic behaviors in order to generate a hypothesis. For Step 2, we asked participants to curate at least 30 input images to test, compared to the approximately five they tested in the earlier workshops. We also scaffolded their input generation by having them fill out a 2D input organizer with a different variable along each axis (see an example in Figure \ref{fig2}). This enabled participants to visually brainstorm and keep track of their input ideas so that these could be systematic (logical) and thorough (covering all the possibilities they could think of). When testing their inputs in Step 3, we gave each group a table for them to keep track of their input images, output images, and notes about changes from inputs to outputs. This table also scaffolded their analytic work in Step 4, as participants drew on both descriptions of changes they noticed in filtered images as well as percentages that summarized trends across their entire corpus. Lastly, we identified a clear audience for their audit reports in Step 5, participants’ peers, the original designers of the filters being audited. We also discussed the objective of the audit: to provide feedback for their peers to improve the design of their filter.

\subsection{Data Collection \& Analysis}
During the summer workshop, three primary sources of data were collected that contributed to the development of this case study: recordings of image and video artifacts youth created (e.g., PDFs of input organizers, audit reports in the form of TikTok videos), and screen recordings of their work on project computers and phones. Out of the four groups that conducted audits of each other's filters, only three were recorded (due to one student not consenting to participation in the study). We began our analysis by reviewing materials from the three consenting groups to select one for the case study. We chose one group of youth, Ishmael and Ziyi, after a primary analysis of artifacts including hypothesis generation notes, input graphs, input-output tables, analysis notes, and audit reports. This review highlighted the robustness of Ishmael and Ziyi’s data set and analysis. Since our aim with this analysis was to demonstrate the possibility of engaging youth with auditing activities, we chose this group as the most promising exemplar.

Once a focal group was determined, we reviewed the computer and phone screen recordings in order to better understand how the artifacts Ishmael and Ziyi created were developed. We watched 15 hours of recordings from the 3 days they spent auditing a peer-designed filter that gave people a red clothes, beginning with the development of multi-modal content logs that documented their primary conversations with screenshots of their work during each of the five steps. Then we rewatched the recordings to produce analytic memos documenting youth’s engagement with each step. From the content logs and analytic memos, the case study \cite{yin2009case} presented below was constructed in order to describe Ishmael and Yizi’s process of auditing a TikTok filter.

\section{Findings}
We present the findings of our case study, focusing on two youths, Ishmael and Ziyi, a 14-year-old male and 15-year-old female, and their respective engagement with the five auditing steps laid out above. Together, they audited a filter designed by their peers. The filter was designed with the intention of giving the user red hair, red clothes, and a background that looked like a cloudy sky. Through their auditing process, Ishmael and Ziyi developed the hypothesis that this filter also enhanced or exaggerated the feminine characteristics of its users. They then designed a set of inputs, ran their tests, and analyzed the outputs, concluding that the filter indeed displayed this bias, reporting that it transformed all input images in their dataset into female-presenting people. 

\subsubsection{Step 1: Developing a Hypothesis}

\begin{figure}[t]
\centering
\includegraphics[width=\columnwidth]{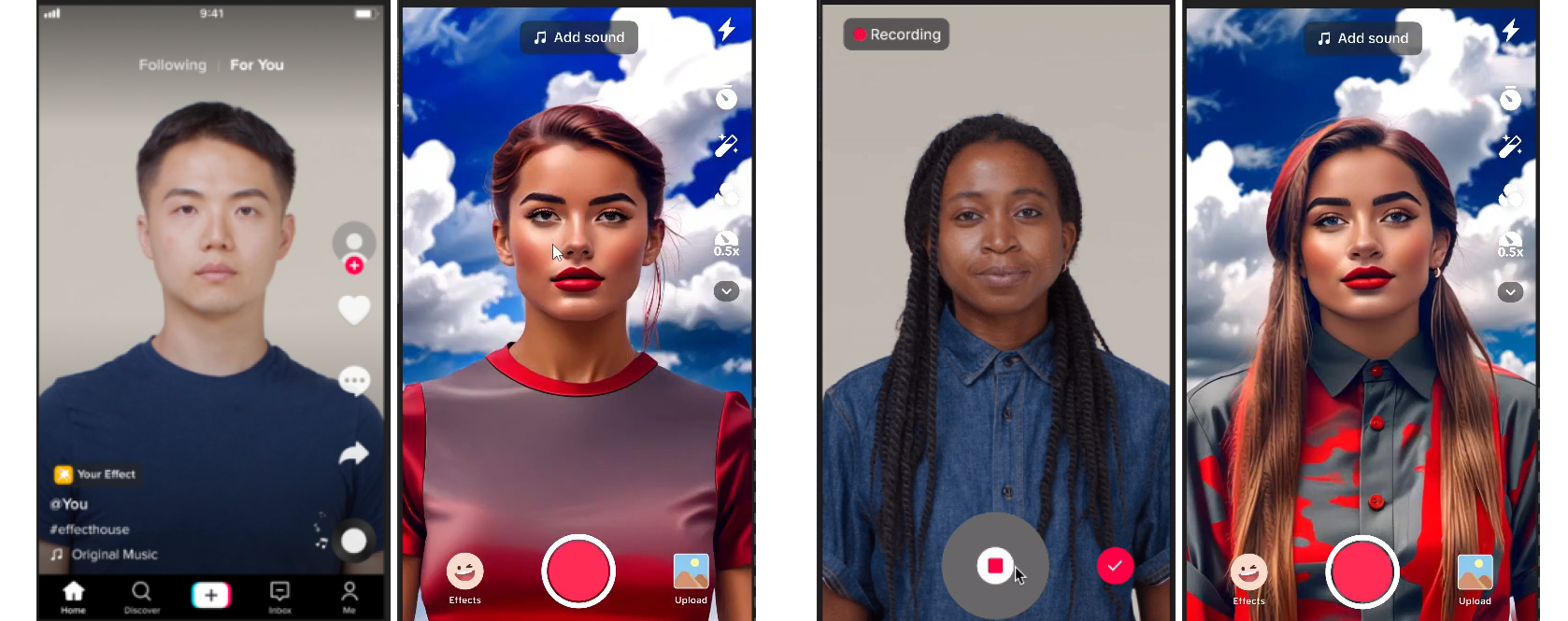} 
\caption{Ziyi tested the filter with the default images available on Effect House. Figure displays two input and output pairs.}
\label{fig1}
\end{figure}

Before developing a hypothesis, Ziyi and Ishmael spent some time playing with the filter by first testing it on the default images available on Effect House (see Figure \ref{fig1}). After a few minutes, Ziyi explained, “I noticed that it turns everyone into a woman, and it makes the hair red.” Ishmael noted that the filter gave everyone thick eyebrows. He loaded the filter on his phone and started testing it on himself, realizing that the filter made him look like a light-skinned woman and gave him red nails and blush on his cheeks. Ziyi experimented with the filter, using it on pictures of herself taken from different angles. After a while, they decided that they wanted to test if the filter always made people female-presenting. Ishmael explained: “It changes the individual’s gender automatically to female, and then, like enhancing, it enhances female characteristics.” Ziyi and Ishmael’s initial open-ended exploration of the filter and its behaviors was instrumental in helping them develop their hypothesis.

\subsubsection{Step 2: Generating a Set of Systematic, Thorough, and Thoughtful Inputs}

\begin{figure}[t]
\centering
\includegraphics[width=\columnwidth]{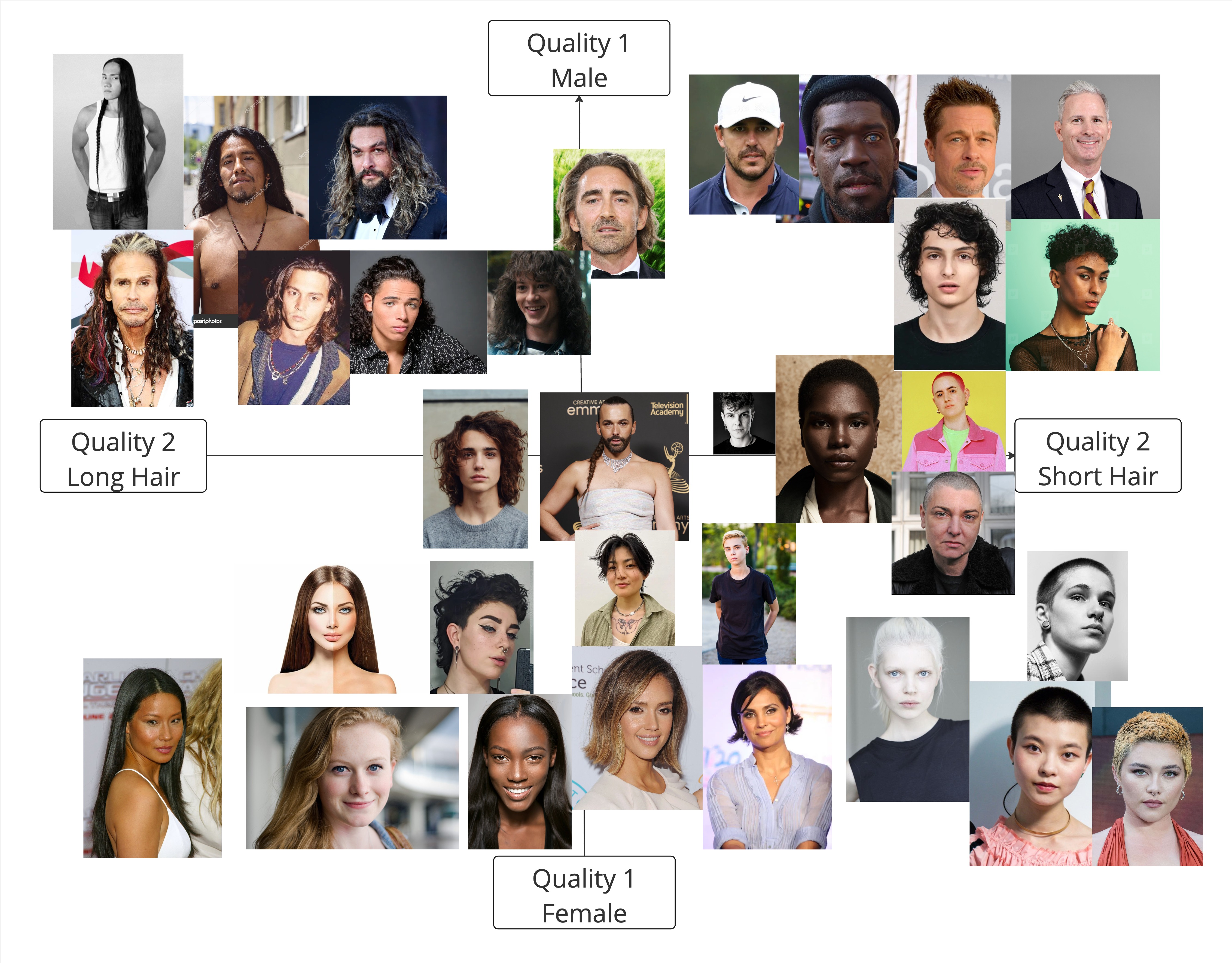} 
\caption{Inputs generated by Ziyi and Ishmael on the two-axis organizer.}
\label{fig2}
\end{figure}

We observed that for youth selecting inputs was one of the most challenging aspects of conducting an audit. Ishmael was clear that they needed images of both male and female people and people with different hair styles. Ziyi also thought it was important to make sure people in the images had different skin tones to observe if the filter worked in the same way for people of different racial backgrounds. This was particularly notable in light of the fact that the group’s hypothesis did not make any direct mention of race or skin color. They started using the organizer to put pictures of male and female celebrities. Ishmael noted that “some of them have to go in the middle [between female and male].” As the group was having this discussion, one of the science center educators, Dolly, approached and asked, “Do you want pictures of only male and female people? Or also non-conforming people?” Ishmael agreed, “Yeah, we also need non-binary people.” Ziyi added, “Maybe we should also think about hair length because it is gender-coded.” They started adding images to their input organizer with hair length and gender as the two axes (see Figure \ref{fig2}), also ensuring they had people of a diverse range of skin colors and racial backgrounds in their dataset. For every image selected, the pair discussed where to locate it in the input organizer. At the end they had 33 images to test the filter.

\subsubsection{Step 3 - Running Tests}

\begin{figure}[t]
\centering
\includegraphics[width=0.8\columnwidth]{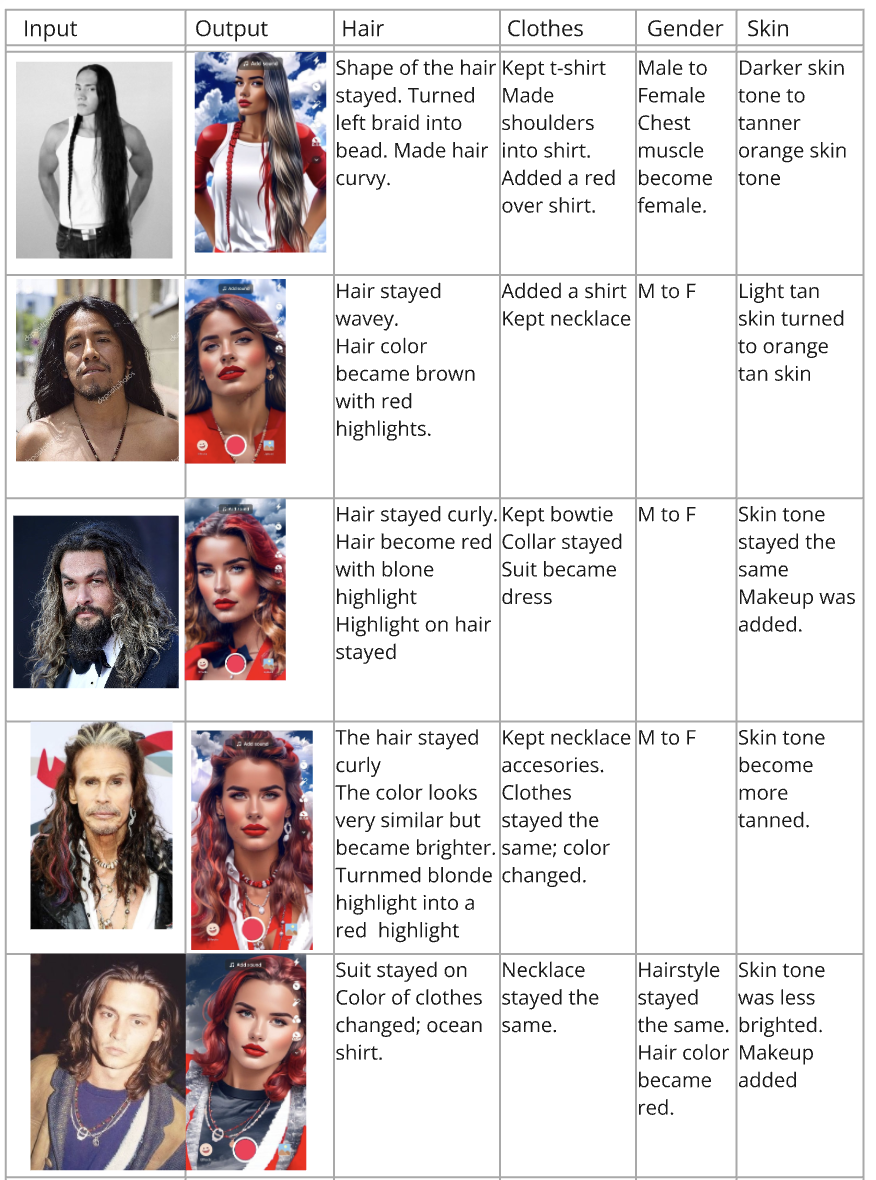} 
\caption{Table with input and output pairs with notes.}
\label{fig3}
\end{figure}

\begin{figure}[t]
\centering
\includegraphics[width=0.9\columnwidth]{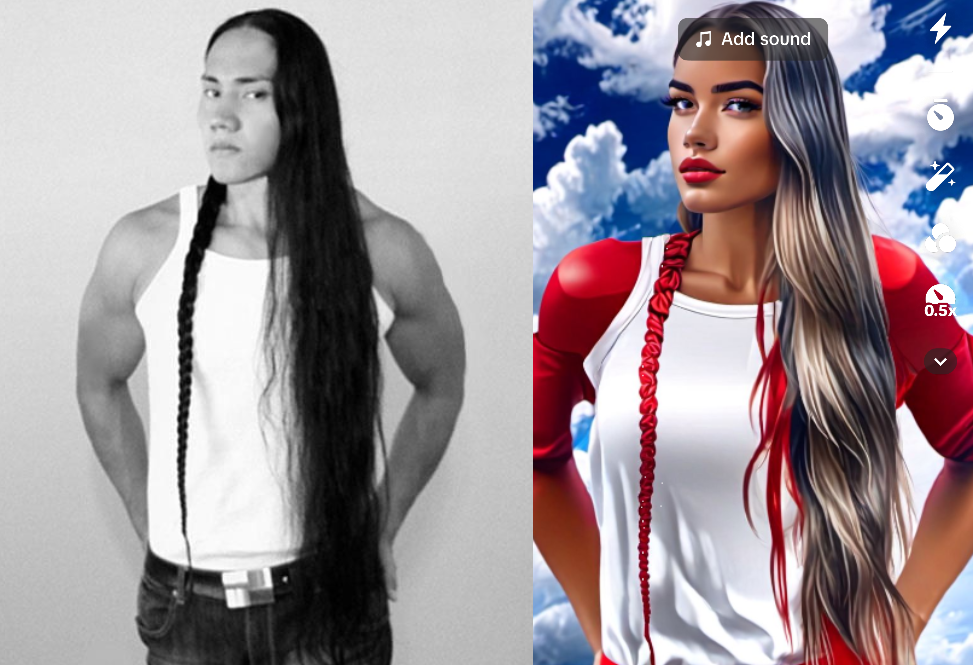} 
\caption{Input-output pair for the first test conducted by Ziyi and Ishmael.}
\label{fig4}
\end{figure}

In the testing phase, Ziyi and Ishmael imported the pictures from their input organizer into Effect House, ran the filter and copied the outputs into a document where they kept track of input and output pairs using a table provided by instructors (Figure \ref{fig3}) . For every image, they discussed their observations and took notes in the table, blending steps 3 (tests) and 4 (analysis) of the auditing process. In taking notes, they first decided on different aspects of the images they wanted to consider . The group agreed to make a note if there were any changes in hairstyle, clothes, gender, or skin tone of the output image. For example, for the first image they tested (see Figure \ref{fig4}) they noted that the hair became more curvy, a braid became a bead, and the clothes changed. The person went from wearing a white tank top with bare arms to a white and red long sleeve shirt. They also noted that the filter changed the gender presentation of the input figure, adding feminine features like blush on their cheeks, red lipstick, and breasts; the skin tone also changed from a “darker skin tone to a tanner orange skin tone” (Ishmael). After annotating two outputs as a team, Ziyi and Ishmael divided the work, with Ziyi running the tests and adding the images to the spreadsheet and Ishmael annotating them. 

\subsubsection{Step 4 - Analyzing the Data}

After conducting the test and qualitatively annotating each of the input-output pairs, Ziyi and Ishamel examined their data as a whole, calculating percentages for gender, hair style, and eurocentric features in the outputs. Ziyi explained: “What we noticed is that it creates a female picture regardless of a person’s gender. We put in a bunch of inputs with a variety of skin tones and gender identities, and we noticed that all of them came out to be female.” They noted that, although their input images were 60\% male, 30\% female, and 30\% non-binary people, all of the outputs looked female. More specifically, they recorded, all of the outputs had blush, tanned skin, and eurocentric features. They also recorded that 26\% of inputs had their hair style changed, and 75\% of the outputs had red hair. 

\subsubsection{Step 5 - Reporting Results}

The group decided to report their audit findings in a TikTok video. Ishmael suggested they make the video humorous, like other TikTok videos he had seen, to engage the audience. To do so, Ishmael changed his voice and introduced the project as an analysis that shows that the filter transformed everyone “into a skibidi\footnote{“Skibidi” is a popular pseudoword among teenagers used as a filler or expletive depending on the context.} female.”  In the video, Ishmael and his friend Brooklyn-May introduced the audit study, sharing the hypothesis and how they decided on different inputs. Then Ziyi reported on the findings from the study. The pair subsequently recorded the video a second time, calling the first one a “blooper.”

\section{Discussion}
This paper provides a first illustration of how we can empower youth to systematically and critically evaluate machine learning applications through algorithm auditing. We proposed a five-step model for learners to engage in algorithm auditing activities and provided a case study in which a team of high school youth used these five steps to conduct an algorithm audit of a TikTok filter. We developed these five steps based on a review of prior literature on expert- and user-driven algorithm auditing. In the following paragraphs, we address some of the affordances and challenges of how the steps scaffolded youths’ engagement and how youth’s practices relate to expert- and user-driven auditing.

Dividing algorithm auditing into different steps scaffolded youth by directing their attention to key features on how to systematically evaluate the performance of the filters. We further scaffolded youth’s engagement with each step by modeling what each step looked like and providing guidance and structured time for youth to conduct their investigations. For example, for step 1, we modeled what a hypothesis could look like and showed examples of hypotheses developed by other youth. In addition, we provided them with time to become familiar with the filter they were auditing through open-ended exploration or play. Here, using the filter on Effect House's default images and on themselves was key for them to begin observing potentially harmful system behaviors that required further investigation. Following, we asked them to document moments when the filters did not work as expected and finally to formulate a hypothesis based on their explorations. Another area that required scaffolding was input generation to support youth in coming up with systematic, thorough, and thoughtful inputs. Introducing the two-axis input organizer (Figure \ref{fig2}.) supported learners in coming up with diverse inputs, but as we saw in the case of Ziyi and Ishamel, learners may consider the diversity of inputs across more than two variables.

Future iterations of auditing activities should further scaffold the creation of inputs,  data analysis, and reporting of findings. Having learners create a large input data set can be a challenge due to the perceived repetitive nature of the task. As we saw in our case study, the team built a dataset of only 33 images to test their hypothesis. Here we see opportunities to explore how generating inputs could be a class-wide group activity where youth build a collaborative dataset or how existing datasets could be leveraged by youth to test their hypotheses. In terms of data analysis, while we asked youth to count and describe patterns they observed in the data, this process could be better scaffolded so that youth can further analyze data by applying statistics knowledge and creating data visualizations. Finally, to report on the findings, youth in our study created TikTok videos, while these short videos were fun to make, other forms of reporting to emphasize computational communication \cite{lui2019student} could be explored, including having youth create infographics, data visualization projects, or longer videos in which they can explain how they conduct an audit and showcase their findings.

Adopting this five-step model for youth required us to think about algorithm auditing from a learner-centered perspective \cite{soloway1994learner}, that is, acknowledging that the goals of learning about auditing are different than those of conducting expert- or user-driven audits. The goal for learners is to understand and experience auditing as a systematic method to evaluate ML-powered systems from the outside-in and not to conduct expert-like audits or user-like one-off observations. Yet, as we observed in the analysis of the case study, incorporating aspects of expert-driven auditing supported youth in being systematic in generating inputs, analyzing data, and drawing conclusions based on evidence. For instance, despite the difficulty of generating all inputs in a single step before beginning to test, we felt this structure was important as it avoided the ad hoc and reactive nature in which everyday users create inputs \cite{devos2022toward}. It also supported youth in engaging, albeit on a smaller scale, with some of the discussions that experts also have when designing protocols to test their hypotheses \cite{metaxa2021auditing}. At the same time, practices observed in user-engaged auditing were particularly helpful to support youth in generating hypotheses. Like everyday users in user-driven audits \cite{shen2021everyday, devos2022toward}, through casual interaction, youth were able to make ad hoc observations and build hypotheses about potentially harmful behaviors. These kinds of ad hoc observations are also present in some expert-driven audits, as expert auditors sometimes draw on their own everyday interactions and user experiences to inspire their hypotheses \cite{buolamwini2018gender,sweeney2013discrimination}. In scaffolding auditing learning activities, we observed how incorporating aspects of expert-driven auditing supports learners in understanding how audits are conducted systematically, while drawing from user-driven auditing practices enabled us to build on youths’ everyday lived experiences as users of AI/ML systems. 

\subsection{Implications \& Future Work}

Next steps for this work involve exploring how the five steps could be used to integrate algorithm auditing activities in classroom settings. This work must  consider a variety of issues, such as choice of AI/ML systems that youth audit, collaborative arrangements, and teacher professional development. To start, in our study youth audited peer-designed TikTok filters. However, Tiktok’s upcoming legal challenges and restricted access to TikTok and Effect House platforms on school devices present challenges for conducting similar work in school settings. To overcome this impending obstacle, future work must identify other applications that could be audited in school settings and that are relevant to youths’ lives, and also identify other viable community spaces in which this particular tool may remain a viable option, such as community centers and after-school programs. 

A second important consideration for implementing algorithm auditing learning activities concerns collaborative arrangements. In our particular case, we decided to have youth work in pairs. As we could observe in our case study, even without the explicit assignment of roles, Ziyi and Ishamel adopted their own work arrangement when running tests and recording observations. While working in small teams was a productive arrangement, we noted that the design of extensive test data remains a challenge. Here future work might investigate how youth in a class or workshop can collectively build a data set, with each student or team contributing a number of test cases, resulting in a more extensive set a single team or student could assemble. 

Finally, we must acknowledge the central role teachers will play in implementing successful algorithm auditing activities in classroom settings. In our case, workshop organizers and staff were instrumental in providing scaffolding and feedback that helped to guide youth through the five steps. We have already started participatory design work with a group of experienced K-12 CS educators to bring algorithm auditing activities into high school classrooms and encourage the proliferation of such efforts. We hope this work can contribute to empowering youth to systematically investigate how ML systems work and what their implications are.

\section{Acknowledgments}
This work was supported by National
Science Foundation (NSF) grant \#2333469. Any opinions, findings, and
conclusions or recommendations expressed in this paper are those
of the authors and do not necessarily reflect the views of NSF,
the University of Pennsylvania, or Columbia University. We thank Lucia Kulzer for support in data collection.

\bibliography{aaai25}

\end{document}